# The $c=\hbar=G=1$ - question

Bernhard Lesche

Physics Department Universidade Federal de Juiz de Fora , Juiz de Fora MG Brazil

e-mail: lesche@fisica.ufjf.br

Abstract: The question whether the invariant speed $c$, Planck constant $\hbar$, and gravitational constant $G$ can be or should be put equal to 1 is analyzed. The discussion is based on fundamental considerations concerning the notion of physical quantity. It is found that the issue is not a matter of appropriate unit selection. Further it is shown that classical space-time geometry and quantum mechanics can be formulated in such a way that the invariant speed and the action quantum are truly one and do not have to be set equal to one.

## 1 Introduction

In a recent publication Michael J. Duff, Lev B. Okun and Gabriele Veneziano exposed three interesting diverging opinions about the question of how many fundamental constants are necessary to formulate and apply the physical laws [1]. The question is whether the invariant speed $c$, Planck constant $\hbar$, and gravitational constant $G$ can be or should be put equal to 1 and what the meaning of equations of type $c=1$ actually is. The divergence of opinions indicates that the issue has not been settled. Therefore we shall address this question again.

Any graduate student who has gone through good quantum mechanics and relativity courses should be able to follow our arguments. We shall start with some very basic questions which should stand at the beginning of any physics course but which are usually handled with little care.

## 2 Physical quantities

A physical quantity is a property of physical objects, and a physical object is an object that can interact directly or indirectly with our senses. But not every property of physical objects is a physical quantity! For instance beauty is not a physical quantity. In order to call a property $Q$ a physical quantity its definition should comply with certain requirements:

a) The set of objects $D_Q$ for which the quantity $Q$ is meaningful should be defined.

b) There should be a rule for deciding whether two objects from $D_Q$ have the same *value* of $Q$.

c) There should be a rule that defines the *sum of values* of $Q$.

Some quantities (but not all) permit also a comparison of values in the sense of larger and smaller.

These requirements need more detailed explanation:

**Comment concerning requirement a):** In general the *domain*, i.e. the set $D_Q$ of admissible objects, is so diversified that it seems difficult to give a general definition.

For instance, if you think of spatial distance we may talk of the distance between two atoms in a molecule, of the distance between two galaxies or of the distance between a car and a bicycle. In order to start with well defined sets one may first take some very restricted set of objects and later extend the quantity at hand to larger domains. For instance, in the case of spatial distance $d$ we may begin with a domain $D_d$ formed of pairs of points marked on some solid body.

**Comment concerning requirement b):** The rule that defines equality of values should be an equivalence relation on the set $D_Q$. An equivalence relation "~" naturally divides a set into classes of equivalent elements. In the case of a physical quantity $Q$ the equivalence classes correspond to the values of the quantity. $\mathcal{A}$ and $\mathcal{B}$ are equivalent iff they have the same value of $Q$.

$$\mathcal{A} \sim \mathcal{B} \quad \Leftrightarrow \quad Q_\mathcal{A} = Q_\mathcal{B} \tag{1}$$

$Q_\mathcal{A}$ is the value of $Q$ that can be attributed to the object $\mathcal{A}$. So if somebody defines a physical quantity he or she has to provide a criterion to decide whether two values are equal. But so far we have not said what a value of a quantity is. We shall come to that point later.

Let us see an example: The quantity *spatial distance*. As mentioned above we may start with a domain that consists of marked pairs of points on a solid body. The equivalence relation can be defined as follows: the pair of points $(A, B)$ is equivalent to the pair $(A', B')$ iff the spikes of a compass that fit into the points $(A, B)$ also fit into the points $(A', B')$. If this is the case we shall write $d_{(A,B)} = d_{(A',B')}$.

**Comment concerning requirement c):** The sum rule of item 3) should allow the determination of a $\mathcal{C} \in D_Q$ for any given $\mathcal{A}$ and $\mathcal{B} \in D_Q$ so that $\mathcal{C}$ represents the value $Q_\mathcal{A} + Q_\mathcal{B}$. But this rule has to comply with certain conditions. First of all it has to be compatible with the equivalence classes. That means the value $Q_\mathcal{A} + Q_\mathcal{B}$ must not depend on the particular class representatives $\mathcal{A}$ and $\mathcal{B}$. If $\mathcal{A}'$ and $\mathcal{B}'$ are objects such that $Q_\mathcal{A} = Q_{\mathcal{A}'}$ and $Q_\mathcal{B} = Q_{\mathcal{B}'}$ then the object $\mathcal{C}'$ that is obtained from $\mathcal{A}'$ and $\mathcal{B}'$ should have the same value as the object $\mathcal{C}$ that one obtains starting with $\mathcal{A}$ and $\mathcal{B}$.

$$Q_\mathcal{A} = Q_{\mathcal{A}'} \land Q_\mathcal{B} = Q_{\mathcal{B}'} \Rightarrow Q_\mathcal{A} + Q_\mathcal{B} = Q_{\mathcal{A}'} + Q_{\mathcal{B}'} \tag{2}$$

Further the rule should fulfill the conditions:

$$Q_\mathcal{A} + Q_\mathcal{B} = Q_\mathcal{B} + Q_\mathcal{A} \tag{3}$$

$$Q_\mathcal{A} + \{Q_\mathcal{B} + Q_\mathcal{C}\} = \{Q_\mathcal{A} + Q_\mathcal{B}\} + Q_\mathcal{C} \tag{4}$$

The requirements (1) - (4) are conditions of self consistency of the quantity $Q$. They seem trivial; however in the laboratories they are never fulfilled. One may consider these conditions as an ideal case and deviations are ascribed to experimental errors.

Generally physical quantities have a zero value "0" that has the property

$$Q_\mathcal{A} + 0 = Q_\mathcal{A} \tag{5}$$

for all $Q_A$. If this value exists it is necessarily unique.

The associative property (4) permits one to forget about the brackets and one may write either side of equation (4) in the simple form $Q_A + Q_B + Q_C$. One may aggregate even more terms in that sum. We shall be especially interested in multiple sums where one single value gets summed several times. These multiple sums of a single value shall be written as a multiple of that value:

$$n Q_A \underset{Def.}{=} \underbrace{Q_A + Q_A + Q_A ... + Q_A}_{n \text{ times}} \tag{6}$$

The majority[1] of physical quantities have the following property: If $n Q_A \neq 0$ one can conclude from an equation $n Q_A = n Q_C$ that $Q_A = Q_C$. This type of quantity we shall call a *linear quantity*. The value $Q_B$ of a linear quantity $Q$ and a positive integer $k$ uniquely determine a value $Q_A$ such that $Q_B = k\, Q_A$. We shall write this value as $Q_A = Q_B / k$:

$$Q_B = k\, Q_A \quad \Leftrightarrow \quad Q_A = \frac{1}{k} Q_B \qquad \text{for any positive integer } k \tag{7}$$

If one combines the multiplications (6) and (7) one gets multiplications with positive rational numbers. The following rules are valid:

$$a\{b Q_A\} = (ab) Q_A \tag{8}$$

$$a Q_A + a Q_B = a\,(Q_A + Q_B) \tag{9}$$

$$a Q_A + b Q_A = (a+b) Q_A \tag{10}$$

Equation (10) allows us to define $0 Q_A \underset{Def.}{=} 0$.

Some physical quantities permit finding for every value $Q_A$ a corresponding value $Q_{***A}$ such that

$$Q_A + Q_{***A} \;=\; 0 \;. \tag{11}$$

$Q_{***A}$ is uniquely determined by $Q_A$. The equations (10) and $0 Q_A = 0$ justify writing this unique value as $(-1) Q_A$. With this operation we extended multiplication to negative rational numbers. An extension to real numbers is mathematically convenient but has no experimental counterpart.

Apart from the sum of values one may define the difference. $Q_A - Q_B$ is defined as the solution of the equation $Q_B + X = Q_A$. For quantities that have the elements $(-1) Q_A$ for all $Q_A$ the uniqueness of $X$ is guaranteed. For quantities without the values $(-1) Q_A$ the uniqueness of $Q_A - Q_B$ has to be required as an additional condition of self consistency.

---

[1] There exist exceptions. For instance the quantity *angle* may be defined in several different ways. One way gives a circular range of values and the implication $n Q_A = n Q_C \;\Rightarrow\; Q_A = Q_C$ does not hold for $n Q_A \neq 0$. Another exception is the pseudo momentum of electrons in a crystal lattice.

For quantities without the values $(-1)Q_\mathcal{A}$ it is convenient to extend the set of values formally to include the values $(-1)Q_\mathcal{A}$. These extensions generally turn out to have physical meaning. For instance the square of the modulus of a vector is a quantity such that there is no element in the domain of this quantity that has the value $-\|\vec{a}\|^2$. However, $-\|\vec{a}\|^2$ is useful: one can define the scalar product of vectors as

$$\vec{a}\cdot\vec{b} \underset{def.}{=} \frac{1}{4}\left\{\|\vec{a}+\vec{b}\|^2 - \|\vec{a}-\vec{b}\|^2\right\}.$$

The set of values of a linear physical quantity $Q$ forms a linear space $V_Q$ or can be extended minimally to a linear space. We shall call this linear space the *value space* of the quantity.

Before we proceed we shall address the question of what a value actually is. One might be tempted to say that the equivalence classes are the values. But this is not always so. See the following counter example: Let $Q$ be some linear physical quantity with a given domain $D_Q$, a given equivalence relation $\sim_Q$ and a given sum rule. The quantity $\hat{Q} = 5Q$ has exactly the same domain, the same equivalence classes and the same sum rule. However, we cannot say that $Q$ and $\hat{Q}$ are the same quantity, because if they were the same we should have the right to write the equation $\hat{Q} = Q$, which obviously implies $5 = 1$. To avoid this one has to select a basic set of quantities. For these basic quantities one may define the values as the equivalence classes. Other *derived* quantities have values that are names attributed to the classes. So let $Q$ be a basic quantity. If $\mathcal{A} \in D_Q$ it belongs to the class $Q_\mathcal{A}$, which is the value of $Q$ attributed to the object $\mathcal{A}$. When one deals with the quantity $\hat{Q}$ one uses the class $5Q_\mathcal{A}$ as a name of the class $Q_\mathcal{A}$ and this name is the value $\hat{Q}_\mathcal{A}$. In general we shall define a value of a quantity to be a name of an equivalence class. In some cases the equivalence classes can be used as their own names. Whatever objects we may choose to be the values of a given quantity the operational rule that defines the sum of values has to be considered as part of the value space.

For a given linear quantity $Q$ one may choose a basis in the value space $V_Q$ so that any particular value of $Q$ can be written as a linear combination of basis values:

$$Q_\mathcal{A} = a_1 Q_1 + a_2 Q_2 + a_3 Q_3 + \ldots\ldots \tag{12}$$

where $a_1$, $a_2$... are numbers and $Q_1$, $Q_2$, ... are elements of the basis. To measure a value of $Q$ means to determine the numbers $a_1$, $a_2$... experimentally. The minimal number of basis values necessary to write any possible value of $Q$ depends on the quantity and this minimal number shall be called its dimension. For instance, spatial distance, mass and speed are one dimensional. Ordinary velocity, acceleration and force are three dimensional. And there exist higher dimensional tensorial quantities. In principle our notion of physical quantity admits even infinite dimensional quantities. For instance the functional relationship of polarization and electric field $\vec{P}(\vec{E})$ is an infinite dimensional quantity that characterizes a dielectric substance. However, the use of such infinite dimensional quantities is not common. People prefer to write $\vec{P}(\vec{E})$ as a Taylor series expansion and characterize the dielectric material by an infinite number of finite dimensional quantities (susceptibility tensors).

In the case of a one dimensional quantity the chosen basis value is called a *unit*. One may choose more than one unit for a given one dimensional quantity. From a theoretical point of view such redundant use of units does not make sense. But for practical reasons several operationally independent unit values may be useful.

It is a widespread mistake to believe that the sum of values of a one dimensional physical quantity can be defined expressing the values in terms of a unit. If you ask your first year physics students how to sum distances you will inevitably get the answer: "Well, I express the distances in cm and then I simply sum them: 2 cm + 5 cm = 7 cm." This is entirely wrong! One needs to know how to sum distances already in order to know what 2 cm means! The manufacturer of the ruler needed the sum rule in order to put the millimeter scratches on the straightedge. In fact, he also needed that sum rule to decide whether the straightedge was really straight. So the definition of sum is prior to the definition of unit!

There is no universal sum rule for all physical quantities. Every single quantity has its own rule. This rule is part of the definition of the quantity. It is instructive to define the sum rule of spatial distances. First one defines an order relation for distances: With the help of a compass one marks all points with a given distance $d$ from a given point $A$. This gives a sphere of radius $d$ and center $A$. Then one defines that the points inside the sphere have a distance from $A$ smaller than $d$ and the ones outside have a distance from $A$ larger than $d$. Once we can compare distances in the sense of larger and smaller we can define the sum as follows: Let $d_1$ and $d_2$ be two distance values (equivalence classes). We choose a pair of points $(A, B)$ in the class $d_1$. Then we collect all points that have the distance $d_2$ from point $B$. Next we chose the point $C$ on that sphere that has the largest distance from $A$. The pair $(A, C)$ is a representative of the class $d_1 + d_2$. An immediate consequence of this definition is the triangle inequality.

3. **The algebra of physical quantities and fundamental constants.**

In the following we shall restrict ourselves to finite dimensional linear quantities. Let $Q$ be a finite dimensional linear quantity with value space $V_Q$. The set of all linear mappings that map $V_Q$ to the real numbers is naturally a linear space, which is called the dual space of $V_Q$ and we shall write it as $V_Q^*$. Generally there exist physical quantities whose values lie in $V_Q^*$. Especially if $Q$ is one dimensional one can define a quantity $1/Q$ with domain $D_Q \setminus \{\mathcal{A} \mid Q_\mathcal{A} = 0\}$ and with values in $V_Q^*$ such that

$$\left(1/Q\right)_\mathcal{A} [Q_\mathcal{A}] = 1 \quad \text{for all } \mathcal{A} \in D_Q \setminus \{\mathcal{A} \mid Q_\mathcal{A} = 0\} \tag{13}$$

In this equation $\left(1/Q\right)_\mathcal{A} [Q_\mathcal{A}]$ means the application of $\left(1/Q\right)_\mathcal{A}$ on the value $Q_\mathcal{A}$. The quantity $1/Q$ is also written as $Q^{-1}$ and it is called the inverse of $Q$. If $Q_\mathcal{U}$ is a unit of the quantity $Q$ the dual basis $Q_{\mathcal{U}}^{-1}$ is a natural unit of the quantity $Q^{-1}$. If "second" (s) is a unit of time then $s^{-1}$ is simply the dual basis of the basis element s.

Let $Q$ and $P$ be two linear quantities with respective domains $D_Q$, $D_P$ and value spaces $V_Q$ and $V_P$ (These value spaces may also be dual spaces of other value spaces). If $D_Q \cap D_P \neq \varnothing$ we can define the tensor product of $Q$ and $P$. This is the quantity $Q \otimes P$ with domain $D_Q \cap D_P$ such that $(Q \otimes P)_{\mathcal{A}} = Q_{\mathcal{A}} \otimes P_{\mathcal{A}}$ for all $\mathcal{A} \in D_Q \cap D_P$. The value space of $Q \otimes P$ is $V_Q \otimes V_P$. If at least one of the quantities is one dimensional all elements of $V_Q \otimes V_P$ can be considered pairs of values. If both quantities are higher dimensional then there exist elements in $V_Q \otimes V_P$ that cannot be considered a pair of values. The requirement $D_Q \cap D_P \neq \varnothing$ is not really a serious problem. If $D_Q \cap D_P = \varnothing$ we remember that the domains can be extended. We may naturally extend the quantities $Q$ and $P$ to the Cartesian product set $D_Q \times D_P$ and these extended quantities can be multiplied.

If the value space of $P$ is the dual space of $V_Q$ then any $Q_{\mathcal{A}} \otimes P_{\mathcal{B}}$ defines a linear mapping $(Q_{\mathcal{A}} \otimes P_{\mathcal{B}}): V_Q \to V_Q$. In the case that $Q$ is also one dimensional this mapping can be written as a multiplication with the number $P_{\mathcal{B}}[Q_{\mathcal{A}}]$. So in the one dimensional case the application of a dual value on a value can be identified with the tensor product of these values. It is custom not to write the tensor product signs in the case of one dimensional quantities. If you see a value of a speed written as 5 m/s in reality you have a tensor product of the value meter with the dual basis of second.

With the tensor products one may define powers of quantities. In the case of one dimensional quantities one can further define non-integer powers that do not have the interpretation of tensor products, but these are not interesting for the present issue, and we shall not go into these details[2].

Now imagine you have a set $O$ of linear quantities with a common domain. We may build all sorts of tensor products and form the direct sum of all these value spaces. This gives a huge linear space $\mathbf{V} = \underset{q \in O}{\oplus} V_Q$ and the tensor product endows this space with the structure of an algebra[3].

One may define new quantities from old ones using the algebraic operations in $\mathbf{V}$. Sometimes this may lead to a quantity $\tilde{Q}$ that defines exactly the same structure on its domain as some previously defined quantity $Q$. The discovery of such coincidence, if unexpected, is celebrated and it is expressed by saying that there exists a fundamental constant $K$. This constant is a linear bijective mapping that maps the new value space onto the old one, $K: V_{\tilde{Q}} \to V_Q$, in such a way that the equivalence classes in the domain that correspond to the values $\tilde{Q}_{\mathcal{A}}$ and $K[\tilde{Q}_{\mathcal{A}}]$ are the same for all $\tilde{Q}_{\mathcal{A}} \in V_{\tilde{Q}}$. Then one writes the physical law:

---

[2] In fact, later we shall use the square root of a quantity that is defined in a value space of type $V \otimes V$. But the definition of this kind of power is obvious.

[3] Due to the presence of dual spaces the set of real numbers $\mathbb{R}$ will also be present in this direct sum.

$$Q = K\left[\tilde{Q}\right] \quad (14)$$

In general, the old and new value spaces $V_Q$ and $V_{\tilde{Q}}$ correspond to operationally different rules of equivalence and sum. Therefore they should, in principle, be considered different spaces. As $Q$ and $\tilde{Q}$ define the same structure on the common domain the presence of both value spaces in the algebra **V** can be considered unnecessary. There are two ways of getting rid of this redundancy: one may simply eliminate the old value space $V_Q$ from the algebra or one may identify $V_Q$ and $V_{\tilde{Q}}$ by declaring that $\tilde{Q}_\mathcal{A}$ and $K\left[\tilde{Q}_\mathcal{A}\right]$ are the same objects for all $\tilde{Q}_\mathcal{A} \in V_{\tilde{Q}}$. The second possibility reduces the fundamental constant $K$ to the identity mapping; $K = \mathbf{1}$. This choice expresses our trust in the physical law (14) to such an extent that we consider irrelevant whether the operational rules of $Q$ or of $\tilde{Q}$ are used to determine the values.

There are thousands of publications with statements of the type "we use natural units such that $c = \hbar = 1$". The identification of value spaces, which corresponds to setting a fundamental constant equal to 1, is not induced by a particular choice of units. No particular choice of basis in the spaces $V_Q$ and $V_{\tilde{Q}}$ will turn these spaces equal. The bijective mapping $K$ is determined by the condition that $\tilde{Q}_\mathcal{A}$ and $K\left[\tilde{Q}_\mathcal{A}\right]$ correspond to the same equivalence classes for all $\tilde{Q}_\mathcal{A} \in V_{\tilde{Q}}$ and not by a choice of bases. The decision to consider $K\left[\tilde{Q}_\mathcal{A}\right] = \tilde{Q}_\mathcal{A}$ is not induced by a choice of basis nor is it related to a choice of basis. We may choose the basis in the original value space $V_Q$ as the image $\left\{K\left[\tilde{Q}_l\right]\right\}_{l=1,..n}$ of a basis $\left\{\tilde{Q}_l\right\}_{l=1,...n}$ without setting $K = \mathbf{1}$ and inversely we may maintain independent choices of basis of $V_Q$ and $V_{\tilde{Q}}$ even setting $K = \mathbf{1}$. In the case of a one dimensional quantity this latter possibility corresponds to the use of two operationally independent units of a single quantity. Later we shall see an example of this situation.

Before considering the cases $c = 1$ and $\hbar = 1$ we shall discuss a classical example of elimination of a redundant value space.

Consider the quantity *angle*. We shall discuss an angle definition that leads to a linear quantity. The domain of that sort of angle is the set of all ordered pairs of half straight lines in a given plane that start at a common point (vertex) together with a winding string. The winding string is a continuous curve in the plane that starts at some point of the first half line, and ends at some point of the second half line and never touches the vertex (compare figure 1). The equivalence classes can be defined by specifying the operations that leave the angle constant: translations, rotations in the plane and continuous deformations of the winding string always keeping the winding string apart from the vertex. A sum of angles is constructed by translating and rotating the second angle object so that both vertexes coincide and that the first half line of the second angle object falls onto the second half line of the first one. Then we rearrange the winding strings so that the second one becomes the continuation of the first one and finally we eliminate the coinciding half lines. Then we get a new angle object that represents the sum of the original angles.

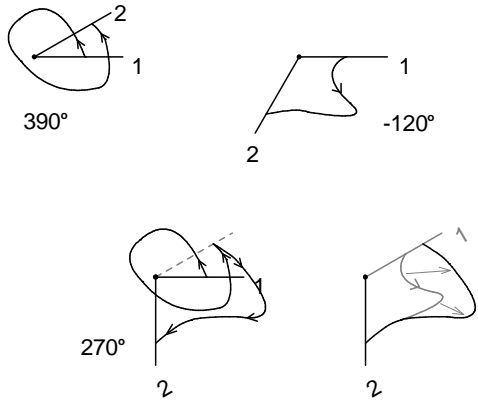

Fig. 1 Angle objects and the determination of a sum of angles.

The elements of the value space are the equivalence classes. Let us call this value space $V_\angle^{class}$. At some stage of history someone had the clever idea to accompany the winding string by a circle with center at the vertex and to use the ratio of the (signed) arc length of that circle and the radius as a measure of angle. Arc length and radius take values in the same value space and therefore the ratio is a pure number. That means we have now a second value space, which is simply $\mathbb{R}$. The numbers can be used as alternative names of the equivalence classes and we have a natural bijective linear mapping from $V_\angle^{class}$ into $\mathbb{R}$. With this second value space one can verify whether two angle objects have the same angle and whether a third angle object has an angle that is the sum of the angles of the other two objects without the use of the original rules. Therefore the original value space $V_\angle^{class}$ can be eliminated. This elimination is not induced by a choice of a unit. It is established by a new definition of the quantity. This new definition is operationally independent of the original one, but it results in the same structure on the set of angle objects.

We remark in passing a curious question: Is the unit "rad" an element of the original value space $V_\angle^{class}$ or is it a number? If it were an element of $V_\angle^{class}$ one would not be allowed to write:

$$\cos\alpha = 1 - \frac{1}{2!}\alpha^2 + \frac{1}{4!}\alpha^4 - \ldots\ldots \quad (15)$$

because a linear combination of different powers of rad would not result in a number. If $\mathrm{rad} \in \mathbb{R}$ then it would just be a funny way of writing the number 1. Sure enough we would like to have equation (15) as a valid formula. So rad = 1. It is in fact superfluous to write this unit. It only serves to remind the reader that here a number is being used as a name of a class of angle objects. Despite the fact that rad = 1, it is a unit. The set of numbers also has units. 1 is a natural unit in $\mathbb{R}$ and there are other numbers that are frequently used as units, for example a dozen or the Loschmidt number $6.02214179(30) \times 10^{23}$. Steradian (sr) is another funny way of writing the number 1. In the next section we shall show that $c$ is still another funny way of writing 1.

**4. Classical Space-Time Geometry**

The human mind has a tremendous predisposition to introduce the notion of space. This is due to the millions of years of evolution of our species on firm ground. An intelligent species that evolved in the oceans far away from the ground and from the shores probably would think directly in terms of events. Doing physics one discovers that space is a mere illusion. If you look at a scratch mark on a metal surface that indicates a point in space you receive the messages of light scattering events and your brain detects an astonishing similarity among these events. It is this similarity that makes our brains believe that a point exists. Once we have understood this we can teach our brains to think in terms of events directly.

Imagine you install four clocks somewhere. These clocks need not be atomic clocks and no specific calibration is needed. You may think of these clocks simple as generators of increasing numbers. Attached to the clocks we have signal receptors and memories. Now an event occurs and some sort of signal is emitted in all directions. The arrivals of this signal at the clocks are registered in the four memories together with the arrival times measured by the respective clocks. This procedure attributes four numbers to the event and this sort of 4-tuple serves as a coordinate system. Apart from events that really got registered we imagine an infinite set of events that would have been registered at certain 4-tuples of coordinate values. Perhaps not all events can be registered by the same system of four clocks and we may be obliged to use several systems. And still other coordinates could have been used. But we shall require that the transformation of coordinates should be differentiable an infinite number of times. So the infinite set of events has the structure of a differentiable manifold. We shall call this manifold Space-Time and denote it by *ST*. The light scattering events of the above mentioned scratch mark would form a curve in that manifold and this sort of curve associated with a small object is called the world line of the object. In the following we shall introduce the structure of space-time geometry using facts that could in principle be verified experimentally, although technically this might be difficult is some cases.

At any event $e \in ST$ one can define a tangent space $T_e$ in the usual way. The tangent vectors are equivalence classes of differentiable curves that start at $e$. The equivalence relation is defined as follows:

$$\alpha \sim \beta \quad \Leftrightarrow \quad \text{for all real differentiable functions } f : \lim_{\lambda \to 0} \frac{1}{\lambda} \{ f(\alpha(\lambda)) - f(\beta(\lambda)) \} = 0 \tag{16}$$

where $\alpha$ and $\beta$ are two differentiable curves at $e$; $\alpha(0) = \beta(0) = e$. The sum of vectors is defined the following way: let $\vec{a}, \vec{b} \in T_e$ be two vectors and $\alpha \in \vec{a}$, $\beta \in \vec{b}$ curves. A curve $\gamma$ that fulfills

$$\lim_{\lambda \to 0} \frac{1}{\lambda} \{ f(\gamma(\lambda)) - f(\alpha(\lambda)) - f(\beta(\lambda)) \} = 0 \tag{17}$$

for all differentiable functions $f : ST \to \mathbb{R}$ is in the class $\vec{a} + \vec{b}$. Multiplication of a vector by a number is very simple: If $\alpha \in \vec{a}$ then the curve $\tilde{\alpha}$ with $\tilde{\alpha}(\lambda) = \alpha(x\lambda)$ is an element of the class $x\vec{a}$. *Tangent vector at event $e$ is a four dimensional physical quantity*, $T_e$ is its value space and any vector in $T_e$ is a value of that quantity.

Next we define a distance between events. Let $e_A$ and $e_B$ be two events. We shall define the *temporal distance* of $e_A$ and $e_B$ with the following procedure: We prepare an atomic clock in such a way that the clock witnesses the events. That means the world line of the clock has to pass through the events. Moreover, we demand that the clock shall not be perturbed by any electromagnetic interaction. Therefore the world line necessarily has to be a world line of a freely falling particle. The modulus of the time difference of the events shown by that clock is the temporal distance of $e_A$ and $e_B$, and we shall write this value as $\tau(e_A, e_B)$. The value space of temporal distances shall be written as $V_\tau$. Unlike the distance of points in Euclidian space, not all pairs of events will have a measurable distance. There exist pairs of events such that no clock can travel between them. The pairs of events that can be connected with a clock shall be called *time-like* pairs.

A differentiable curve $\alpha$ in *ST* shall be called time-like if the pair of events $\langle\alpha(\lambda_1),\alpha(\lambda_2)\rangle$ is time-like for all parameter values $\lambda_1$, $\lambda_2$. For a time-like curve with endpoints $\alpha(\lambda_A)$ and $\alpha(\lambda_\Omega)$ one defines the temporal length as the infimum of the sum of temporal distances of partitions of the curve (compare figure 2).

$$\tau(\alpha;\lambda_A,\lambda_\Omega) \underset{def.}{=} \inf_{\text{all partitions}} \sum_k \tau\big(\alpha(\lambda_k),\alpha(\lambda_{k+1})\big) \qquad (18)^4$$

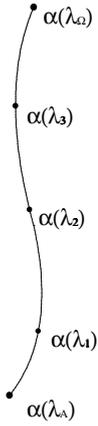

Fig. 2 Partition of a time-like curve.

This definition is motivated by the experimental fact (twin paradox) that a refinement of a partition can only diminish the value of the sum of temporal distances (temporal distances obey an inverse triangle inequality).

The proper time of a time-like curve $\alpha$ is defined as

$$\tau_\alpha(\lambda) \underset{def.}{=} \begin{cases} \tau(\alpha;0,\lambda) & \text{if } \alpha(\lambda) \text{ later than } \alpha(0) \\ -\tau(\alpha;0,\lambda) & \text{if } \alpha(0) \text{ later than } \alpha(\lambda) \end{cases} \qquad (19)$$

In this definition we used the fact that time-like events can be compared in the sense of earlier and later.

A tangent vector $\vec{a} \in T_e$ shall be called time-like if there exist time-like curves in the class $\vec{a}$. The time-like vectors form the union of two convex cones $P_e$ and $F_e$ (past and future) with vertex in 0 and the vectors in $P_e$ are obtained from the vectors in $F_e$ by multiplying with $(-1)$. The vectors on the border[5] of the cones are called null-like[6] and all others space-like. For time-like vectors $\vec{a}$ one finds experimentally that for all time-like curves $\alpha \in \vec{a}$ the derivatives of proper time at $\lambda = 0$ are equal. The value of that derivative is a property of the vector and can be used to define a new quantity:

$$q(\vec{a}) \underset{def.}{=} \left[\frac{d\tau_\alpha(\lambda)}{d\lambda}\bigg|_{\lambda=0}\right]^2 \qquad \text{with} \quad \alpha \in \vec{a}, \ \alpha \text{ time-like} \qquad (20)$$

The values of this quantity lie in the value space $V_\tau \otimes V_\tau$. We shall now extend this new quantity to other vectors. First let us consider a space-like vector $\vec{s}$. One can find time-like vectors $\vec{a}, \vec{b} \in F_e$ such that

$$\vec{s} = \vec{b} - \vec{a} \qquad (21)$$

As $\vec{a}, \vec{b} \in F_e$ and $F_e$ is a convex cone one has also $\vec{a} + \vec{b} \in F_e$ and the value $q(\vec{a}+\vec{b})$ is well defined. Then experimentally we discover the following astonishing facts:

$$2q(\vec{a}) + 2q(\vec{b}) - q(\vec{a}+\vec{b}) \quad \text{does not depend on the choice of } \vec{a} \text{ and } \vec{b} \qquad (22)$$

---

[4] In Euclidian space the length of a path would be defined with a supremum.

[5] The topology of $\mathbb{R}^4$ induces a topology in *ST* and this induces a topology in the tangent space.

[6] Usually these vectors are called light-like. But we shall avoid this term in order to make clear that we do not assume any particular behavior of light signals.

and
$$2q(\vec{a})+2q(\vec{b})-q(\vec{a}+\vec{b}) \;<\; 0 \qquad (23)^7$$

The statement (22) means that we have discovered a property of the vector $\vec{s}$. We shall define

$$q(\vec{s}) \underset{def.}{=} 2q(\vec{a})+2q(\vec{b})-q(\vec{a}+\vec{b}) \qquad (24)$$

As $q(\vec{a})>0$ for time-like vectors the inequality (23) motivates the definition

$$q(\vec{n}) \underset{def.}{=} 0 \qquad \text{for } \vec{n} \text{ null-like} \qquad (25)$$

Then two more astonishing experimental facts can be verified:

$$q: T_e \to V_\tau \otimes V_\tau \qquad \text{is continuous} \qquad (26)$$

and

$$q(\vec{a}+\vec{b}) + q(\vec{a}-\vec{b}) \;=\; 2q(\vec{a}) + 2q(\vec{b}) \qquad \text{for all } \vec{a},\vec{b}\in T_e \qquad (27)$$

Now we can define the scalar product of two tangent vectors:

$$\vec{a}\cdot\vec{b} \underset{def.}{=} \frac{1}{4}\{q(\vec{a}+\vec{b})-q(\vec{a}-\vec{b})\} \qquad (28)$$

**Theorem:** The scalar product is a bilinear mapping from $T_e \times T_e$ to the value space of squares of time $V_\tau \otimes V_\tau$.

**Proof:**
Let $\vec{a},\vec{b},\vec{c}\in T_e$ be given vectors. We shall show that

$$\vec{a}\cdot(\vec{b}+\vec{c}) \;=\; \vec{a}\cdot\vec{b}+\vec{a}\cdot\vec{c} \qquad (29)$$

Using the definition (28) we get

$$\vec{a}\cdot(\vec{b}+\vec{c})-\vec{a}\cdot\vec{b}-\vec{a}\cdot\vec{c} \;=\;$$
$$=\frac{1}{4}\{q(\vec{a}+\vec{b}+\vec{c})-q(\vec{a}-\vec{b}-\vec{c})-q(\vec{a}+\vec{b})+q(\vec{a}-\vec{b})-q(\vec{a}+\vec{c})+q(\vec{a}-\vec{c})\} \qquad (30)$$

Adding and subtracting $q(\vec{a}+\vec{b}-\vec{c})$ inside the curly brackets and applying equation (27) several times one sees that the right hand side is in fact zero.

Applying this result for the case that $\vec{c}=\pm n\vec{b},\; n\in\mathbb{N}$ one concludes readily that

$$\vec{a}\cdot(x\vec{b})=x\,\vec{a}\cdot\vec{b} \qquad (31)$$

for all rational $x$. Continuity permits extending this to $\mathbb{R}$. Equations (29) and (31) mean that the product is linear in the right hand factor. It is obviously symmetric and hence bilinear. ∎

---

[7] There is a natural order relation on the value space $V_\tau \otimes V_\tau$ such that $\tau\otimes\tau>0$ for all $\tau\in V_\tau,\; \tau\neq 0$.

The scalar product is a tensor; it is an element of the space $V_\tau \otimes V_\tau \otimes T_e^* \otimes T_e^*$. This tensor is called the metric of space-time. Using the dual basis of some coordinate basis $\{\partial_\mu\}_{\mu=1,..4}$ it can be written as $\cdot = g_{\mu\nu} dx^\mu \otimes dx^\nu$ and one has

$$\vec{a} \cdot \vec{b} \;=\; g_{\mu\nu} \, dx^\mu [\vec{a}] dx^\nu [\vec{b}] \tag{32},$$

where $g_{\mu\nu} = (\partial_\mu) \cdot (\partial_\nu) \in V_\tau \otimes V_\tau$. A Riemann connection compatible with this metric can be defined and also the curvature tensor.

The essential point is that the geometric structure of space-time can be introduced without defining the value space of spatial distance. If one now introduces spaces into the theory in order to help the human minds the notion of spatial distance will get automatically the value space $V_\tau$. A point in space would be a world line of these light scattering events that happen at a scratch mark. The distance of two neighboring points representing world lines can be defined as the length of a shortest curve that connects these world lines going from one line to the other always staying orthogonal to the point representing world lines. In order to define the length of such curves one needs the notion of length of a space-like vector. The length of a space-like vector is defined as

$$\|\vec{s}\| \underset{def.}{=} \sqrt{-\vec{s} \cdot \vec{s}} \qquad \text{for } \vec{s} \text{ space-like} \tag{33}$$

and the length of a time-like vector

$$\|\vec{a}\| \underset{def.}{=} \sqrt{\vec{a} \cdot \vec{a}} \qquad \text{for } \vec{a} \text{ time-like} \tag{34}$$

Let $\vec{b} \in T_e$ be a vector and let $\vec{a}$ be a time-like vector. We can find a space-like vector $\vec{s}$ orthogonal to $\vec{a}$ such that $\vec{b} = (\vec{s} + \vec{a})\lambda$ with some $\lambda \in \mathbb{R}$. The speed of $\vec{b}$ relative to $\vec{a}$ is the quotient $\|\vec{s}\|/\|\vec{a}\|$. It is a pure number. Now consider the case of a null-like vector $\vec{n}$. This vector can be written as a sum of a space-like vector $\vec{s}$ and a time-like vector $\vec{a}$ orthogonal to $\vec{s}$; $\vec{n} = \vec{s} + \vec{a}$ with $\vec{a} \cdot \vec{s} = 0$. From $q(\vec{n}) = \vec{n} \cdot \vec{n} = 0$ it follows $\|\vec{a}\| = \|\vec{s}\|$ and the null speed is 1 and does not have to be put equal to 1. This value is independent of the choice of $\vec{a}$. This fact is called the invariance of the null speed.

The interesting point is that here $c$ has never been set equal to 1. We formulated the geometry of space-time without the value space of spatial distance $V_d$. The constant $c$ can be a fundamental constant that maps the value space of times $V_\tau$ onto the old value space $V_d$. Setting this fundamental constant equal to 1 corresponds to the decision to consider two different operational rules of distance measurement as equivalent and this decision could be questioned by the dimensionality arguments used by Lev Okun [1]. In the present formulation of space-time geometry the redundant value space $V_d$ has not been identified with $V_\tau$ but it has been cut out from the theory. Speeds can be defined in space-time, they are pure numbers and the null speed happens to have the value 1.

## 5. Quantum Mechanics

Quantum physics reveals some subtle details of the physical quantity concept. One classifies physical objects and considers some to be different states of the same physical system. A physical system is associated with a certain set of observables that can be measured with this system. So for instance, a system of two particles permits measuring two positions, two velocities, energy, and so on. The set of all these observables classifies the system to be a two particle system.

Now classically one assumes that all observables of a system have well defined values and these values depend on the state of the system. The state of the system is experimentally determined by some experimental procedure that furnishes the system. We shall refer to this procedure as the *preparation of the state*. To be realistic one has to admit that real experimental preparation procedures may be imprecise which leads to doubts concerning the values of observables. But this is a mere matter of experimental imperfection. In quantum mechanics the situation is totally different. For a given state $\mathcal{S}$ (even a perfectly well prepared one) only a certain subset $O(\mathcal{S})$ of observables has definite values. One might be inclined then to abandon the distinction of system and state. But this distinction is still useful. This is so because the other observables that do not pertain to the set $O(\mathcal{S})$ can still be measured. That means it is technically possible to perform these measurements. But these measurements ask "wrong questions" because the system with the state $\mathcal{S}$ is not in the domain of the measured quantity. If one asks a wrong question the system will respond with a lie. One can notice that the answer was a lie by repeating all the experimental procedure. Repetitions will lead to different answers. This is not the fault of the system; we forced the system to tell lies by asking inadequate questions. We should have asked a question taken from the set $O(\mathcal{S})$.

One can still extract relevant information from the fluctuating answers to inadequate questions if one registers the relative frequencies of long series of equally performed measurements.

The adequate way of representing this situation is the well known Hilbert space fashion. So for instance the perfectly prepared states can be represented by a normalized state vector and yes–no questions by closed subspaces. If the state vector pertains to such a subspace and we ask the corresponding yes-no question the system will definitely answer "yes", if it pertains to the orthogonal complement the answer is definitely "no" and if it does not pertain to either of theses two spaces the question is inadequate and the probabilities of a positive answer are given by the square of the norm of the projection of the state vector. Imperfectly prepared states can be represented by a density operator and observables by essentially self adjoined operators. As our present issue concerns the question of units and the nature of values we have to be still more careful and represent observables by elements of the tensor product of our value algebra $\mathbf{V} = \underset{q \in O}{\oplus} V_Q$ and the space $\mathcal{S}(\mathcal{H})$ of essentially self adjoined operators on a complex Hilbert space $\mathcal{H}$. All this can be expressed by the *first law of quantum physics*:

There exist mappings $R$ and $\rho$ that map observables of a given system on elements in the tensor product $\mathcal{S}(\mathcal{H}) \otimes \mathbf{V}$ and states on self adjoined trace class operators with trace 1 such that for all states $\mathcal{S}$ and all observables $\mathcal{O}$ the mean values $\overline{\mathcal{O}_\mathcal{S}}$ in very long series of equally performed experiments become equal to $Tr(R(\mathcal{O})\rho(\mathcal{S}))$.

The pair of mappings $\langle R, \rho \rangle$ shall be called a *representation of the quantum system*. We called the above statement the first law of quantum physics because it is very similar to the first and second law of thermodynamics. These laws state the existence of the state functions $U$ and $S$ requiring general properties of these functions, but they do not provide the functions for a given system. For any system these functions have to be determined experimentally or calculated with the help of statistical mechanics. The first law of quantum physics also expresses the mere existence of a representation.

The problem of constructing a representation for a given system is the real challenge in quantum mechanics. We may say that the solution of the *representation problem* contains all relevant physics of a system (including the solution of the dynamics). Textbooks offer an unsatisfactory solution of the representation problem. This pseudo solution is known as *quantization*. The quantization maps observables of an imagined classical system onto operators in a Hilbert space. This is not a solution of the representation problem because classical electrons do not exist. At best certain states exist such that the electron behaves like a classical particle.

A clean treatment of quantum mechanics should solve the representation problem without any classical observables. A solution of the representation problem of a given system has to start with the physical characterization of observables and states. The only known way to characterize observables physically is with the help of *cinematic symmetries*. A cinematic symmetry of a quantum system is a mapping that substitutes old states with new states and old observables with new ones in such a way that all mean values in long series of equal experiments remain unchanged.

$$S : \mathcal{S} \mapsto \tilde{\mathcal{S}} \quad \text{and} \quad S : \mathcal{O} \mapsto \tilde{\mathcal{O}}$$
$$\text{such that} \quad \overline{\mathcal{O}_\mathcal{S}} = \overline{\tilde{\mathcal{O}}_{\tilde{\mathcal{S}}}} \quad \text{for all } \mathcal{O} \text{ and } \mathcal{S} \tag{35}$$

Once we find a cinematic symmetry we can investigate the behavior of experimental outcomes when we submit an observable to this substitution without substituting the state. The behavior of outcomes can be used to characterize the observable.[8]

Imagine we do have a representation of a quantum system. Then any substitution of old observables with new ones and old states with new states will induce Hilbert space substitutions of old operators with new ones. The condition that the mean values should remain unchanged together with the representation requirement $\overline{\mathcal{O}_\mathcal{S}} = Tr(R(\mathcal{O})\rho(\mathcal{S}))$ implies that the operator substitutions have to conserve the traces $Tr(R(\mathcal{O})\rho(\mathcal{S}))$. In cases without super selection rules Wigner´s theorem [2] tells us that the operator substitutions are necessarily of the form

$$A \quad \rightarrow \quad \tilde{A} = UAU^{-1} \tag{36}$$

where $U$ is a unitary or anti-unitary operator. Now the main symmetry operations available for our construction act on some basic observables as simple space-time transformations. On other observables they may act in a complicated way. Let us see what that means. Let $\mathcal{O}$ be an observable from that basic set. This observable corresponds to some concrete measuring apparatus and process, which is characterized by an inner structure (documented by technical designs, circuits, computer codes and the like) and space-time markers such as scratch marks on the apparatus and event markers that characterize the beginning of processes. The whole measuring process occurs in some large space-time region $M$. Now let $s$ be a bijective mapping of space-time into space-time that conserves the geometry of $M$. Then we shall say that the symmetry $S$ acts as the space-time symmetry $s$ on the observable $\mathcal{O}$ if the old space-time markers $m_1, m_2, \ldots$ get replaced by the markers $s(m_1), s(m_2), \ldots$ and the inner structure of the apparatus and process remains unchanged. The requirement of unchanged inner structure restricts the choice of mapping $s$ to geometry conserving mappings at least in the region $M$ of interest. The whole construction of a representation is a long story and we shall concentrate here only on the few aspects that bare direct relevance on our question at hand.

---

[8] A cinematic symmetry need not be symmetry of the Hamiltonian. Therefore the method is quite general.

Now imagine we have found a set of cinematic symmetries that act on some basic observables as space-time symmetries and these space-time symmetries form a Lie-group. For instance think of rotations around a given axes or space-time translations. We shall suppose that the relevant space-time region $M$ is sufficiently flat so that all tangent spaces $T_e$ can be identified with a single space $T$, whose vectors may be taken to be equivalence classes of pairs of events. For an infinitesimal rotation angle $\delta\varphi$ and an infinitesimal translation $\delta\vec{t}$ we may write the unitary operators that represent these symmetries in Hilbert space as

$$U_{\delta\varphi} = \mathbf{1} - i\, K[\delta\varphi] \tag{37}$$

and

$$U_{\delta\vec{t}} = \mathbf{1} - i\, \bar{P}[\delta\vec{t}] \tag{38}$$

where $K[\delta\varphi]$ and $\bar{P}[\delta\vec{t}]$ are self adjoined or essentially self adjoined operators that depend on $\delta\varphi$ and $\delta\vec{t}$ respectively. Due to the infinitesimal character of $\delta\varphi$ and $\delta\vec{t}$ these dependences are linear. So in the case of equation (37) one can write $K[\delta\varphi]$ in the form of a product $K[\delta\varphi] = J\,\delta\varphi$ of a self adjoined operator $J$ and the angle $\delta\varphi$ (of course we use the rad-quantification of angles). In the case of translation, $\bar{P}[\bullet]$ is naturally an element of the space $\mathcal{S}(\mathcal{H}) \otimes T^*$. Iterating the infinitesimal operations an infinite number of times, one gets the operators that represent finite rotations and finite translations:

$$U_\varphi = \exp\{-i\varphi J\} \tag{39}$$

and

$$U_{\vec{t}} = \exp\{-i\bar{P}[\vec{t}]\} \tag{40}$$

$\varphi$ is the rotation angle and $\vec{t} \in T$ a translation vector.

Now we can solve part of the representation problem by going in the opposite direction; from Hilbert space to the laboratory. We may ask whether we can identify these mathematical objects with objects in the laboratory. We shall do so in a general form, writing generically

$$U_\lambda = \exp\{-i\lambda G\} \tag{41}$$

where $\lambda$ is the group parameter and $G$ is an operator (generator). We are looking for an observable $\mathcal{G}$ whose representative is the operator $G$. One can recognize the eigenstates of that observable by subjecting them to the symmetry operation. They should be invariant with respect to these operations. So suppose one managed to find a lot of these eigenstates $g_1, g_2, g_3, \ldots$ .[9] Let us see whether we can determine the eigenvalues experimentally. To that end one looks for a pure state $S$ such that the experimental question " is your state $g_1$ ?" gets a positive answer a fraction $\alpha$ of times $(0 < \alpha < 1)$ and the question " is your state $g_2$ ?" gets a positive answer a fraction $(1-\alpha)$ of times. Then we know that $S$ is a coherent superposition of the states $g_1$ and $g_2$. If one subjects an arbitrary[10] observable $A$ from the basic set to the symmetry operation $A \mapsto A(\lambda)$ and if one measures $A(\lambda)$ with the state $S$ then one finds that the mean values $\overline{A(\lambda)_S}$ oscillate:

$$\overline{A(\lambda)_S} = A_0 + B\cos\left((g_1 - g_2)\lambda + const.\right) \tag{42}$$

where $g_1$ and $g_2$ are the eigenvalues of the eigenstates $g_1$ and $g_2$. Because of the presence of an undetermined phase constant in equation (42) this sort of quantum interference experiment can only determine the absolute value $|g_1 - g_2|$. So we see how absolute values of eigenvalue differences can be measured. In this general form of experiment we cannot expect more than that for the following reason: The representation of a quantum system is not unique. A given representation can always be substituted by another one that differs from the first one by a unitary or anti-unitary transformation. This ambiguity introduces ambiguities of the generators $G$. A unitary transformation may add a constant to a generator and anti-unitary transformations will also change the sign of $G$. In special cases these ambiguities can be removed. The anti-unitary changes can be eliminated by some definite choice during the construction of the representation and the arbitrary constant can sometimes be eliminated depending on the structure of the symmetry group. So, for instance, the structure of the rotation group permits eliminating the arbitrary constants of angular momentum (generators of rotations). Non relativistic linear momentum (generators of space-translations) can also be fixed but non-relativistic energy (generator of time translations) remains with an arbitrary constant. The Lorenz group permits fixing the constants of energy-momentum (generators of space-time translations). So we saw how the generators can be related to observables.

Now $J$ of equation (39) is related to an observable that we shall call angular momentum. The eigenvalues of $J$ are pure numbers. In the case of rotations the function $f(\lambda) = \overline{A(\lambda)_S}$ of equation (42) has to be $2\pi$-periodic. Therefore the smallest difference of eigenvalues of angular momentum is 1. $\overline{P}[\bullet]$ of equation (40) is related to an observable that we shall call energy-momentum. $\overline{P}[\bullet]$ can be written with the help of the dual basis of a coordinate basis:

$$\overline{P}[\bullet] = P_\mu dx^\mu \tag{43}$$

---

[9] In the case of a continuous spectrum one can only find approximate eigenstates. So, for instance, in the case of translation symmetry one can find states that seem invariant under translations $\vec{t}$ when observed with measurement procedures in the region $M$ as long as $\vec{t}$ is small compared to $M$.

[10] This observable should not be invariant with respect to the symmetry operation.

The application of a dual basis element $dx^\mu$ on a translation vector results in a pure number. Therefore the coefficients $P_\mu$ are elements of $\mathcal{S}(\mathcal{H})$ and their eigenvalues are pure numbers. The mass is defined by the relation

$$m^2 = g^{\mu\nu} P_\mu P_\nu \tag{44}$$

where $g^{\mu\nu}$ is the inverse of the metric:

$$g^{\alpha\nu} g_{\nu\beta} = \delta^\alpha_\beta \tag{45}$$

The $g_{\nu\beta}$ are elements of the value space $V_\tau \otimes V_\tau$ and consequently $g^{\alpha\nu} \in V_\tau^* \otimes V_\tau^*$. Then one concludes that the value space of mass is $V_\tau^*$. So $s^{-1}$ is a mass unit. No independent mass value space had to be introduced. We came to that result not by putting $\hbar = 1$. This constant simply is not needed to formulate quantum mechanics. $\hbar$ is the fundamental constant that maps the dual space of times onto the classical mass value space. A formulation of quantum mechanics without quantization has no place for such a mapping. On the other hand, $\hbar$ could also be considered an action unit. Action is a phase angle and hence a pure number. The action unit appears as the smallest eigenvalue difference of angular momentum and it turns out to be 1.

For practical reasons it may be convenient to reintroduce the old classical mass value space together with its unit kg. Then the fundamental constant $\hbar$ is reintroduced and it is a matter of taste whether one sets it equal to one or not. Independent of this choice we have then two independent mass units; the kg and $s^{-1}$ (or the image of $s^{-1}$ under the mapping $\hbar$). This is the actual situation imposed by the international system of units. The simultaneous presence of two operationally independent mass units is justified by the technical difficulty of measuring the Compton wavelength of a macroscopic mass. However, it is possible to relate macroscopic masses indirectly to quantum interference experiments. An attempt to build an experimental quantum-kg has been performed [3]. When the quality of this type of experiment gets further improved the classical kg might be condemned to retirement in a museum.

## 6. Gravity

Quantum mechanics formulated in a flat classical space-time region eliminates the classical value space of inertial mass. If we trust in the equivalence of inertial mass and gravitational mass we may eliminate the only remaining one dimensional value space $V_\tau$ together with its dual space $V_\tau^*$. One may quantify masses by the square root of the ratio of Schwarzschild radius and Compton wave length. This allows one to identify the mass value space $V_\tau^*$ with the numbers very much the same way one can identify the original value space of angles $V_\measuredangle^{class}$ with $\mathbb{R}$. But this construction uses the time value space $V_\tau$ in an intermediate step. Anyway, no choices of units remain and the really important pure numbers that a fundamental theory should be able to explain get clearly exposed.

Unfortunately this mixture of classical space time geometry, quantum mechanics, and gravity, including the gravitational field produced by the quantum system, is inconsistent [4]. We have no unique and convincing quantum gravity. Experiments that might select among possible proposals are still rather limited [5]. In the present article we certainly shall not attempt to solve this problem. But we may see whether the arguments used can give hints.

Some assumptions made in section 4 (Classical Space-Time Geometry) have to suffer radical alterations when quantum gravity is concerned. Which assumption is the most suspicious one? There is a clear answer to that question: the assumption that space-time is a differentiable manifold. First of all the registered coordinate values will have experimental imprecision orders of magnitude larger than the relevant resolution in quantum gravity. Second, the sentence "Apart from events that really got registered we imagine an infinite set of events that would have been registered at certain 4-tuples of coordinate values" clearly pertains to the classical world. From a quantum point of view a physical process that has been registered in four memories is quite a different thing from a process that could have happened and could have been registered. In ordinary quantum field theory space-time points are used as labels for the dynamical variables of the theory. This permits describing the symmetries and a causality condition in a convenient way:

$$U_s \Phi_k(x) U_s^{-1} = \Lambda_{sk}{}^l \Phi_l(s^{-1}x) \qquad (46)$$

$$\left[\Phi_k(x), \Phi_m(x')\right]_\mp = 0 \qquad x, x' \text{ space-like} \qquad (47)$$

But a different kind of label might serve the same purpose. On the other hand, we can be pretty sure that paths are necessary to form gauge-invariant expressions that might represent observables. S. Mandelstam formulated a proposal of a quantum theory of gravity based on paths [6]. The paths are registered by their tangent vectors in a single linear space. Classically this unique space would be the tangent spaces parallel transported along the path up to the initial point. Whether two paths end at the same point depends on the curvature and in quantum gravity curvature will exhibit quantum fluctuations. Therefore the events of our classical manifold will disappear. We recognized that space was an illusion. Probably we shall have to admit that space-time is an illusion too.

**7. Conclusions**

The notions physical quantity, value space, unit and fundamental constant were analyzed. A fundamental constant appears when two operationally independent quantities define the same structure on their common domain. Then the number of chosen units can be reduced. Further one may eliminate a redundant value space. This can be done in two ways: one may formulate a theory without the unwanted space or one may identify spaces setting a fundamental constant equal to 1. The first possibility was exemplified with Classical Space-Time Geometry and with Quantum Mechanics.